\documentclass[sigconf]{acmart}
\setcopyright{none}
\renewcommand\footnotetextcopyrightpermission[1]{}
\usepackage{stfloats} 
\usepackage{listings}
\usepackage{algorithmic}
\usepackage{graphicx}
\usepackage{textcomp}
\usepackage{xcolor}
\usepackage{makecell}
\usepackage{caption}

\lstdefinelanguage{mlir}{
    alsodigit = {.},
    keywords = {stencil.apply, stencil.access, arith.constant, arith.addf, arith.mulf, stencil.return, f32, f64, stencil.index,stencil.temp, hlfir.declare, fir.load, arith.cmpi, fir.if, fir.result, scf.if, scf.yield, func.return, fir.alloca, hlfir.assign, memref.alloca, memref.store, memref.load, memref.alloc, arith.subi, arith.addi, linalg.yield, linalg.reduce, xrtw.num_devices, xrtw.init, xrtw.allocate_buffer, xrtw.buffer_map, xrtw.buffer_sync, xrtw.run, xrtw.wait, scf.for, aie.core, aie.tile, aie.device, aie.objectfifo, aie.objectfifo.acquire, aie.objectfifo.release, aie.objectfifo.subview.access, aie.end}
}

\AtBeginDocument{%
  }

\setcopyright{acmlicensed}
\copyrightyear{2025}
\acmYear{2025}
\acmDOI{XXXXXXX.XXXXXXX}

\acmConference[ISFPGA '25]{ International Symposium on Field-Programmable Gate Arrays}{Feb 27-- Mar 01,
  2026}{Monterey, CA}
\acmISBN{978-1-4503-XXXX-X/18/06}

\begin{document}

\title{Seamless acceleration of Fortran intrinsics via AMD AI engines}

\author{Nick Brown}
\email{n.brown@epcc.ed.ac.uk}
\orcid{0000-0003-2925-7275}
\affiliation{%
  \institution{EPCC at the University of Edinburgh}
  \city{Edinburgh}  
  \country{UK}
}

\author{Gabriel Rodriguez-Canal}
\affiliation{%
  \institution{EPCC at the University of Edinburgh}
  \city{Edinburgh}  
  \country{UK}
}


\begin{abstract}
A major challenge that the HPC community faces is how to continue delivering the performance demanded by scientific programmers, whilst meeting an increased emphasis on sustainable operations. Specialised architectures, such as FPGAs and AMD's AI Engines (AIEs), have been demonstrated to provide significant energy efficiency advantages, however a major challenge is that to most effectively program these architectures requires significant expertise and investment of time which is a major blocker.

Fortran in the lingua franca of scientific computing, and in this paper we explore automatically accelerating Fortran intrinsics via the AIEs in AMD's Ryzen AI CPU. Leveraging the open source Flang compiler and MLIR ecosystem, we describe an approach that lowers the MLIR linear algebra dialect to AMD's AIE dialects, and demonstrate that for suitable workloads the AIEs can provide significant performance advantages over the CPU without any code modifications required by the programmer. 
\end{abstract}



\maketitle

\pagestyle{plain}

\section{Introduction}

Whilst High Performance Computing, HPC, currently relies on mainstream CPUs and GPUs, driven by the incesent demand from scientific computing for increased performance, but at the same time an increased emphasis on energy efficiency, there is increasing interest in leveraging new more specialised hardware technologies. AMD Xilinx's AI engines are one example of this, initially released as part of the Versal Adaptive SoC these engines are vector arithmetic accelerators. AI Engines, or AIEs and we use these two terms interchangeably throughout this paper, adopt a Very Long Instruction Word (VLIW) design and contain a dedicated 512 bit vector unit. There are up to 400 engines on a Versal, and it has been demonstrated that this large amount of raw compute is suitable for a range of High Performance Computing (HPC) workloads \cite{brown2023exploring} \cite{zhang2023new} \cite{klaisoongnoen2024evaluating}.

In 2023 AMD released their Ryzen AI 7000 series of CPUs which combines traditional x86 CPU processing cores with a Neural Processing Unit (NPU). An NPU is an array of second generation AIEs, known as AIE-ML, which are targeted more towards AI workloads. However, this very close coupling of the AIEs with the CPU cores offers a range of potential opportunities for leveraging these specialised engines to accelerate computational kernels, and furthermore the Ryzen AI CPU makes running on AIEs a much more attractive proposition for end users compared to buying a Versal. 

However, a major challenge with AIEs is in the programming of this technology. Developers must use a specialised API and possess a detailed knowledge of the architecture. Moreover, a significant time investment is required to port codes to the AIEs and optimise for performance. This is especially challenging given that kernels are written in C++ or Python, whereas over 60\% of scientific computing workloads on a typical supercomputer are in Fortran \cite{rodriguez2023fortran}.

It is our belief that, in order to drive adoption of AIEs, offloading from the CPU to the NPU must be entirely transparent to the programmer, requiring no effort on their behalf. In this paper we describe an approach which enables the seamless offloading and acceleration of specific Fortran intrinsic calls via AMD's AI engines. This paper is structured as follows; in Section \ref{sec:bg} we explore the background to this work, before in Section \ref{sec:accelerate} describing our approach to offloading Fortran intrinsics on the AIEs. This is then followed by a performance comparison against running these intrinsics on the CPU in Section \ref{sec:performance}, before drawing conclusions and discussing further work in Section \ref{sec:conclusions}.

The contributions of this paper are as follows:
\begin{itemize}
  \item To the best our our knowledge, the first demonstration that scientific programmers can transparently offload parts of their existing codes onto AIEs without any modifications.
  \item Description of a  compilation approach where support for new AIE kernels can be trivially integrated.
  \item A performance exploration of the AMD Ryzen AI NPU for common Fortran intrinsics, demonstrating the NPU provides benefits for specific types of intrinsic.
\end{itemize}

\section{Background and related work}
\label{sec:bg}


AMD, formerly Xilinx, developed their AI Engines to harden commonly used arithmetic operations and combined this with the flexibility of their reconfigurable fabric in the Versal. AIEs follow a Very Long Instruction Word (VLIW) approach where, each cycle, they are capable of issuing up to seven instructions, and an AIE can handle both scalar and vector operations, with the vector unit of size 512 bits. AIEs are arranged in a 2D array, with engines connected to their neighbours in both dimensions, and are able to access the memory from their north, south and west neighbours directly. Furthermore, each engine has four data movers comprising two 32 bit input streams and two 32 bit output streams.

A natural evolution for this technology was to integrate AIEs directly into the CPU, where the CPU contains a Neural Processing Unit (NPU) which is an array of 20 AIEs arranged in five columns of four rows. Each row also contains a memory tile, and four of the columns have an interface tile which connects it to the CPU. The AI engines in the Ryzen AI NPU are the AIE-ML series which, in contrast to the AIEs in the Versal, have been more heavily optimised for AI workloads. Consequently, the AIE-ML contains double the data memory per AIE, 64KB and, unlike the Versal's AIE array, there are five dedicated memory tiles, one per column. Each of these contains 512KB of memory which is decomposed across 16 banks. Each memory tile is equipped with 12 data movers providing a total bandwidth of up to 30 GB/s \cite{aie-ml}. This enhancement to the memory contained within the AIE array provides additional flexibility and a wider set of workloads that can be supported.

AMD have however also removed some arithmetic support in the AIE-ML compared to the first generation AIE, for example the vector units in the AIE-ML series do not natively support int32 or float32 data types in hardware, although these can be emulated. Instead, bfloat16 is provided and int32 integer arithmetic is only natively supported when mixed with int16. Whilst this is a limitation for HPC workloads, the AI engine technology is evolving rapidly and coupling of x86 CPU cores with the NPU is very promising. Consequently, it is still interesting to explore this technology in the context of HPC as future versions could provide increased precision if there is a market demand.

The code that runs on the AIEs comprises of two parts, kernels which are mapped to the AI engines themselves and a graph description which connects kernels and memories together via streams. Kernels operate following the consumer-producer model, where they consume input data from a maximum of two streams and produce results on a maximum of two output streams. These streams can be connected to the CPU directly via an interface tile, to a memory tile or to another AI engine compute tile.

There have been some efforts to address programmer productivity on the AIEs, for instance in \cite{levental2024end} the authors presented an end to end programming model for AIEs and a Python embedded Domain Specific Language (DSL). In \cite{kalkhof2024enabling} the HPX programming framework was enhanced to support AI engines by extending the TaPaSCo framework enabling TaPaSCo FPGA and AIE tasks to be transparently integrated into HPX applications. TaPaSCo \cite{heinz2021tapasco} is an open source toolchain that provides a scriptable flow for the construction of dataflow designs on FPGAs, and an APIs that provides task parallel computing on FPGAs. This tool was enhanced \cite{heinz2024tapas} to also support programming AIEs on the Versal and integrates with the existing FPGA approach. However, all these tools share the same limitation that programmers must learn new technologies and then manually port their codes, potentially requiring rewriting in new languages. By contrast, our view is that programmers must be able to leverage these technologies without any changes necessary to their existing codes or specialised knowledge on their behalf.

\subsection{LLVM and MLIR}

LLVM \cite{lattner2004llvm} provides reusable compiler and tool chain technologies the enable the development of compilers across different languages and hardware. There are many language frontends provided by LLVM and support for a range of hardware backends, with these connected via LLVM-IR. Consequently, an LLVM frontend such as Clang, which provides C and C++, that generates LLVM-IR is able to target any backend, and support for a wide range of architectures including CPUs, GPUs, and FPGAs has been developed. However, LLVM-IR is low level, and it requires significant work by each frontend to target LLVM-IR and results in duplication of compilation infrastructure between frontends.

MLIR, which was first developed by Google and then released open source in 2019, aims to address this issue of duplication by providing many IR dialects and transformations between them. Consequently, instead of targeting the low level LLVM-IR, frontends can translate to a mix of higher level intermediate representations and then leverage existing transformations within MLIR to lower to LLVM-IR. The IR follows a Static Single Assignment (SSA) form, and one of the major strengths of MLIR is that dialects can be mixed and manipulated separately, enabling progressive lowering of the abstraction level ultimately to LLVM-IR.  Because this involves existing dialects and transformations, the MLIR approach enables a much greater sharing of compiler infrastructure between frontends, significantly reducing the overall software effort in developing compilers. MLIR also provides a framework for defining bespoke dialects and transformations. 

MLIR is a sub-project of LLVM, and there are many IR dialects provided as standard including \emph{memref} for memory management and data access, \emph{func} to represent functions and calling between them, and \emph{linalg} which express linear algebra operations. All of these ultimately lower to the \emph{llvm} MLIR dialect, from which LLVM-IR is generated by the \emph{mlir-translate} tool. A considerable community has grown up around MLIR with involvement from many vendors. AMD have invested heavily in this technology with their own fork of MLIR which targets the AIEs. AMD developed several dialects, including \emph{aie} that for streaming connections between AIE compute tiles and direct memory access, \emph{aievec} for vector arithmetic operations, and \emph{adf} to express AMD's Adaptive Data Flow (ADF) graph that connects tiles. For each of these, transformations and optimisations have been developed which ultimately results in a set of instructions that execute across the AIE array.

\subsubsection{xDSL}

One of the disadvantages of MLIR is that programmers must initially learn a range of complex LLVM concepts, and then work with the Tablegen format to describe dialects and keep track of the fast evolving MLIR repository. By contrast, xDSL \cite{xdsl} is a Python based compiler design toolkit which is 1-1 compatible with MLIR. Providing the majority of standard MLIR dialects, as well as numerous additional experimental ones too, these are all expressed in the IRDL \cite{fehr2022irdl} format within Python classes. xDSL enables rapid exploration and prototyping of MLIR concepts, and once these are matured and proven they can then be contributed into the main MLIR codebase  more easily. Because xDSL is 1-1 compatible with MLIR, one is able to arbitrarily go between these technologies during compilation. We have used xDSL to develop the work described in this paper.

\subsection{Flang}

Flang \cite{flang} is LLVM's Fortran frontend and built on-top of MLIR. It began in 2020 as a ground-up rewrite of the previous Flang Fortran compiler, classic Flang, and is now an official component of LLVM. Whilst the objective of Flang is to support the full range of standard Fortran, including being able to adapt to future versions of the language, at the time of writing support for Fortran at or beyond 2003 is still work in progress although Flang is developing rapidly.

\begin{figure}[htb]
\centering
 \includegraphics[width=\columnwidth]{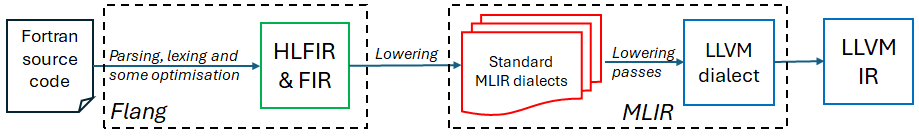}
\caption{Illustration of MLIR-based Fortran compilation flow developed by \cite{brown2024fully} and based upon Flang to generate LLVM-IR.}	
\label{fig:flang-flow}
\end{figure}

Figure \ref{fig:flang-flow} illustrates a sketch of the overarching Flang compilation flow from \cite{brown2024fully} where, after lexing and parsing of a user's Fortran code, some optimisations are undertaken on the AST which is then lowered to Flang's High Level Fortran IR (HLFIR) \cite{hlfir} and Fortran IR (FIR) \cite{fir} dialects. These two MLIR dialects are part of Flang and specifically represent Fortran constructs and concepts. However, Flang is isolated from much of the rest of the MLIR ecosystem and the compiler then directly generates LLVM-IR from these two dialects. In this paper we leverage the work undertaken in \cite{brown2024fully} where, instead of directly generating LLVM-IR from these two dialects, a transformation pass lowers to the standard MLIR dialects and then relies on the rest of the MLIR ecosystem to progressively lower and optimise the IR to generate the LLVM-IR. Not only does this approach enable integration with a wide range of existing infrastructure which is developed by the large MLIR community, moreover the IR can be intercepted at any point and specialised for specific architectures, in this paper for the AMD AIE. 

In this paper we focus on Fortran intrinsics, which are built in procedures defined by the Fortran standard to provide utility functionality. Given Fortran's lineage in scientific computing, a range of intrinsics are defined that undertake calculations and examples include the \emph{sum} procedure which sums all numbers in an array. Whilst the Flang compiler maps these directly to function calls in the Flang runtime library, the work undertaken in \cite{brown2024fully} instead maps these to the linear algebra, \emph{linalg}, dialect. Operations in this \emph{linalg} dialect are then lowered using the existing MLIR infrastructure, for instance to be optimised for the CPU, and the rich source of information about the linear algebra operation can be leveraged to target other architectures, in this work AMD's AIEs.

\section{Seamless offload of Fortran intrinsics to AIEs}
\label{sec:accelerate}
\begin{figure*}[htb]
\centering
 \includegraphics[width=\textwidth]{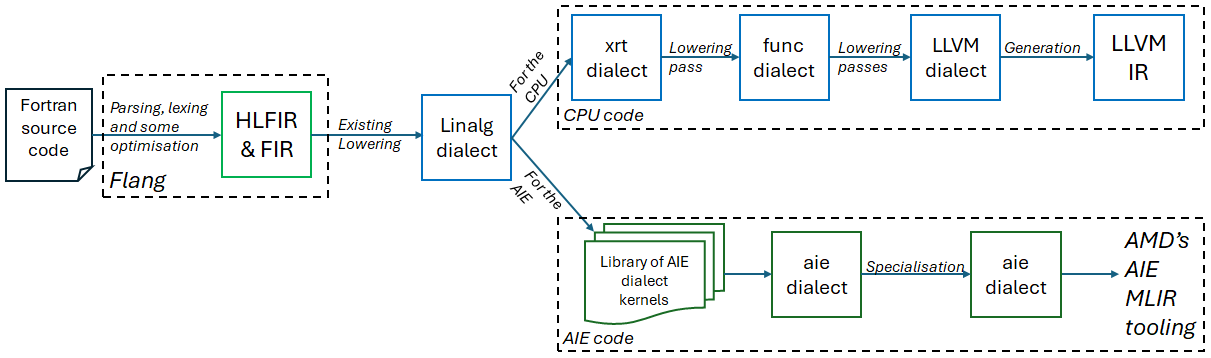}
\caption{Illustration of our overarching compiler approach that offloads selected linear algebra operations to the AIE array}	
\label{fig:flang-aie}
\end{figure*}

To seamlessly offload Fortran intrinsics to the AIEs we have developed transformations and lowerings for operations in the linear algebra, \emph{linalg}, dialect to target the NPU in the Ryzen AI 7000 CPU. Figure \ref{fig:flang-aie} illustrates our approach, which builds upon the compilation flow illustrated in Figure \ref{fig:flang-flow}, where the resulting IR containing  the HLFIR and FIR dialects are lowered to standard MLIR dialects, including \emph{linalg} for intrinsics, at which point we intercept it.

We developed a transformation that analyses operations in the \emph{linalg} dialect and categorises these according to the functionality being undertaken. This is trivial where there is a direct mapping from the intrinsic to the \emph{linalg} dialect's operation, for example matrix multiplication is represented as \emph{linalg.matmul}. However other intrinsics are represented indirectly, for example the \emph{sum} intrinsic results in the \emph{linalg.reduce} operation, and this IR is illustrated in Listing \ref{lst:linalg-reduce} for accumulating over a one dimensional array. The body of the operation at lines 3 and 4 in Listing \ref{lst:linalg-reduce} operate across each element, with a specific element held in \%32 and the running total in \%33. Consequently, where appropriate, our transformation interrogates the body of linear algebra operations, such as a reduction, and categorises it accordingly.

\begin{lstlisting}[language=mlir, frame=lines, label=lst:linalg-reduce, numbers=left, caption=Sketch of the IR corresponding to the Fortran sum intrinsic which uses the \emph{linalg.reduce} operation]
linalg.reduce ins(%29:memref<?xi32>) outs(%30:memref<i32>) dimensions = [0] 
  (%32 : i32, %33 : i32) {
    %34 = arith.addi %32, %33 : i32
    linalg.yield %34 : i32
}
\end{lstlisting}

Once the specific type of linear algebra operation has been identified then, as illustrated in Figure \ref{fig:flang-aie}, our flow generates two components; the CPU side IR to drive the NPU and the AIE side IR that runs on the AIEs in the NPU. The CPU uses AMD's Xilinx RunTime (XRT) to manage and interact with the AIE array, and to aid this we developed an \emph{xrt\_wrapper} MLIR dialect because there is no existing MLIR dialect that drives the AIEs from the CPU. In addition to the dialect, we also developed a transformation that lowers this to function calls, calling into the corresponding runtime library functions on the CPU via MLIR's \emph{func.call} operation.

However integration with XRT was more complex than initially assumed due to name mangling of C++ function and object names, making it difficult to directly call these from the IR. Whilst XRT provides a C interface this is incomplete, for instance it omits the \emph{register\_xclbin} function which is required to set up the AIE array. However, the C++ and C APIs are incompatible and-so as this registration function on the device object in C++ must be called, this then requires the rest of the C++ API to be used. Consequently, we developed a runtime wrapper which is precompiled and linked against on the CPU. This provides a simplified C style interface to XRT that is straightforward to call from the IR, and internally leverages the XRT C++ functions. 

\begin{lstlisting}[language=mlir, frame=lines, label=lst:xrt-dialect, numbers=left, caption=Summary of our XRT wrapper \emph{xrtw} MLIR dialect]
xrtw.num_devices() -> (i32, i1)
xrtw.init(%xclbin_name, %instr_name, %device) : (str, str, i32) -> i1 
xrtw.allocate_buffer(%size, %flags, %group) : (i32, i32, i32) -> (i32, i1)
xrtw.buffer_map(%idx) : (i32) -> (llvm.ptr, i1)
xrtw.buffer_sync(%idx, %direction) : (i32, i32) -> i1
xrtw.run(%b0, ..., %bn) : (i32, ..., i32) -> i1
xrtw.wait() -> i1
\end{lstlisting}

It is this runtime wrapper that our \emph{xrt\_wrapper} MLIR dialect is based upon, and the IR is lowered to calling functions of, with the main operations in this dialect sketched in Listing \ref{lst:xrt-dialect}. The \emph{xrtw.num\_devices} returns the number of NPUs present and is used at runtime to determine whether to launch the instrinsic on the AIE array or instead if none is present to run on the CPU. When buffers are allocated via \emph {xrt.allocate\_buffer} they are assigned an integer identifier which is used as a reference to that buffer in other operations. An example is \emph{xrt.run} which is varadic, accepting any number of buffers which are then passed to the AIE kernel. In MLIR operations can return multiple values, and all operations in our dialect return at-least a boolean, \emph{i1}, which represents success or failure. Whilst this currently only covers a subset of XRT functionality, it is sufficient for driving the NPU in this work and can be extended in the future if required. Once the \emph{xrt\_wrapper} dialect is lowered to function calls, this IR is then provided to standard MLIR transformations, ultimately being lowered to the LLVM-IR MLIR dialect, which is provided to the \emph{mlir-translate} tool that then generates LLVM-IR which is compiled into an object file by Clang.

The second  component is the IR generated from the \emph{linalg} operation that will run on the NPU and undertakes the actual computation. We ported AMD's AI engine MLIR dialects into xDSL to enable us to represent IR in these dialects using that tool. The approach we have taken is to build a library of IR, based upon common linear algebra operations, and the appropriate IR can then be loaded during compilation by the compiler and specialised. We followed this approach because, for a specific linear algebra operation, the IR is very similar and only requires small modifications for each instantiation. Consequently, we pre-generate templates of the MLIR IR which corresponds to each linear algebra operation, load this into xDSL's IRDL format and then store this externally in a library by serialising the IR using pickle. Whilst it is possible to manually write the MLIR IRs for each linear algebra operation, AMD provides a Python interface to their MLIR dialects and a tool that generates MLIR from this, along with a range of numerical examples \cite{mlir-aie}. 

Consequently we store these existing examples, and our own, within this library. The appropriate IR is then loaded during compilation via pickle and deserialised. This IR can be considered a template, and the \emph{specialisation} transformation in Figure \ref{fig:flang-aie} then manipulates this IR to specialise it for each instantiation, for example, by replacing placeholders in the IR with the type of data and number of elements that are being computed with. The resulting IR is illustrated in Listing \ref{lst:air-graph}, for a scalar addition running on a single AIE compute tile which adds one to each input value. FIFOs are created at lines 4 and 5 to link the interface and compute tiles, and then the \emph{aie.core} operation at line 6 specifies the code that will run on an individual AIE compute tile. The FIFOs are configured to be of size 32, int32, elements, and the computation running on the AIEs works in batches of 32 elements. In each batch, between lines 10 and 23, the input and output FIFOs are acquired and then between lines 16 and 21 the input value is loaded from the FIFO, used in the calculation, and stored to the output FIFO. These FIFOs are released at lines 22 and 23, with the compute core looping around and continuing. For brevity, the IR which issues DMA memory copies between the host and the interface tile has been omitted.

\begin{lstlisting}[language=mlir, frame=lines, label=lst:air-graph, numbers=left, caption=Sketch of IR using the AIE dialects that will run on the AIE array for a scalar add example]
aie.device(npu1_1col) {
    %i_tile = aie.tile(0, 0)    
    %comp_tile = aie.tile(0, 2)
    aie.objectfifo @in0(%i_tile, {%comp_tile}, 2 : i32) : !aie.objectfifo<memref<32xi32>>        
    aie.objectfifo @out0(%comp_tile, {%i_tile}, 2 : i32) : !aie.objectfifo<memref<32xi32>>        
    %comp_core = aie.core(%comp_tile) {
      %c0 = arith.constant 0 : index
      %problem_size = arith.constant 32 : index      
      scf.for %arg0 = %c0 to %problem_size {
        %0 = aie.objectfifo.acquire @in0(Consume, 1) : !aie.objectfifosubview<memref<32xi32>>
        %1 = aie.objectfifo.subview.access %0[0] : !aie.objectfifosubview<memref<32xi32>> -> memref<32xi32>
        %2 = aie.objectfifo.acquire @out0(Produce, 1) : !aie.objectfifosubview<memref<32xi32>>
        %3 = aie.objectfifo.subview.access %2[0] : !aie.objectfifosubview<memref<32xi32>> -> memref<32xi32>
        %c0_0 = arith.constant 0 : index
        %c32 = arith.constant 32 : index        
        scf.for %arg1 = %c0_0 to %c32 {
          %4 = memref.load %1[%arg1] : memref<32xi32>
          %c1_i32 = arith.constant 1 : i32
          %5 = arith.addi %4, %c1_i32 : i32
          memref.store %5, %3[%arg1] : memref<32xi32>
        }
        aie.objectfifo.release @in0(Consume, 1)
        aie.objectfifo.release @out0(Produce, 1)
      }
      aie.end
    }
    ...
}
\end{lstlisting}

In the example of Listing \ref{lst:air-graph} the entirety of a core's computation is held directly in the IR, whereas it is also possible to call into external functions via the \emph{func.call} operation. This is useful for more complicated kernels where one can, for example, develop a C++ kernel and compile it into an object file via the chess compiler. Different versions of these external functions, for instance handling distinct data types, can be called and this is materialised during the specialisation step. In this manner, our compilation flow is able to leverage a wide range of existing AIE kernels, with it possible to develop new implementations and optimisations which can then be integrated with our compiler via the dialect library. 

This IR is then fed into AMD's AIE MLIR tooling, which generates the resulting \emph{xclbin} and instruction files that are then launched on the NPU by the CPU code. Our compilation flow also maintains the CPU implementation of the linear algebra operation, for instance leveraging this instead if the data type or data size is not appropriate to run on the NPU.  

\section{Results and evaluation}
\label{sec:performance}
In the experiments reported throughout this section we run on a Ryzen AI 7940HS CPU equipped with 32GB of DRAM. This CPU contains 20 AIE-MLs, arranged in five columns each of four rows. Each column also comprises a memory tile and four of the columns contain an interface tile connecting to the CPU. When running across the AIE array we only leverage the four columns that contain an interface tile, hence we run across 16 AIEs unless otherwise stated. We use GCC version 13.2, Vitis version 2023.1, XRT version 2.18.0, Flang version 20.0.0 (based upon LLVM release 18.1.8), and the latest AIE-MLIR at the time of writing. All results are averaged over ten runs and AIE execution times include the overhead of transferring data to and from the NPU. All CPU code is compiled at optimisation level three, and CPU comparison executables are from Flang using the linear algebra based CPU lowering in \cite{brown2024fully}.

\begin{lstlisting}[language=fortran, frame=lines, label=lst:ftn-sum, numbers=left, caption=Example use of the Fortran sum intrinsic]
integer :: data(100000), result, i
do i=1, 100000
    data(i)=i
end do
result=sum(data)
\end{lstlisting}

We undertook a performance comparison for reduction based Fortran intrinsics, such as \emph{sum} which accumulates the values held in an array. Listing \ref{lst:ftn-sum} sketches the programmer's Fortran code for the \emph{sum} intrinsic, where the \emph{data} array is defined and initialised, and then the intrinsic accumulates its values and returns this in the \emph{result} variable. In this example it is the \emph{sum} intrinsic call that is offloaded to the NPU by our approach.

We developed an AIE implementation for these reduction intrinsics which runs over four columns of the AIE array. A quarter of the array is held in each of the four memory tiles, and then the four compute tiles in that column each consume a quarter of their memory tile's data. The output of each AIE is a single summed number, and these 16 numbers are then read back by the CPU and summed together. Each AIE undertakes a vectorised reduction, operating on vectors of length 512 bits, and AIEs operate in batches of 16384 elements to decouple the data that is being processed from the 64KB of memory available on each compute tile.

Table \ref{tab:basic_intrinsics} reports runtime performance of these reduction intrinsics on the CPU and NPU across data types. \emph{maxval} and \emph{minval} calculate the maximum and minimum array value. bfloat16 is not defined by the Fortran standard, so we developed a transformation to replace float32 in the IR with MLIR's bfloat16 type for those configurations. We also include a \emph{conv-bfloat16} result in Table \ref{tab:basic_intrinsics} for the NPU, where at runtime our XRT wrapper converts between float32 and bfloat16, with float32 on the CPU and bfloat16 running on the NPU.

\begin{table*}[t]
    \begin{center}     
    \begin{tabular}{|c|cccc|ccccc|}
    \hline     
      \textbf{Intrinsic} & \multicolumn{4}{c|}{\makecell{\textbf{CPU runtime (us)}}} & \multicolumn{5}{c|}{\makecell{\textbf{NPU runtime (us)}}}\\
      & \textbf{int16} & \textbf{int32} & \textbf{bfloat16} & \textbf{float32}& \textbf{int16} & \textbf{int32} & \textbf{bfloat16} & \textbf{float32} & \textbf{conv-bfloat16}\\
      \hline
    sum & 606 & 296 & 5187 & 962 & 3107 & 3207 & 3534 & 3511 & 8623 \\
    product & 627 & 305 & 4925 & 1021 & 3111 & 3215 & 3533 & 3464 & 8592 \\
    maxval & 260 & 261 & 286 & 334 & 3214 & 3243 & 3113 & 3115 & 8298 \\
    minval & 265 & 254 & 273 & 355 & 3233 & 3147 & 3261 & 3116 & 8341 \\
    \hline
    \end{tabular}
    \caption{Runtime performance (in microseconds) of reduction based Fortran intrinsics on the CPU and NPU's AIE array, operating on a one dimensional array of size 262144 elements.}
    \label{tab:basic_intrinsics}
    \end{center}
    \vspace*{-\baselineskip}
\end{table*}

It can be seen from Table \ref{tab:basic_intrinsics} that, apart from the bfloat16 data type, the CPU very significantly outperforms the NPU at all data type configurations. The CPU performs poorly with bfloat16, most likely because it is being emulated in software. It can be seen that converting between bfloat16 and float32 in our XRT wrapper adds significant overhead. When exploring why the NPU was so much slower than the CPU we found that the majority of the runtime is constant regardless of the problem size and when run is first called significant setup time is incurred. 

We therefore repeated our experiments, calling the intrinsic operation on the NPU twice, ignoring the initial time and recording the runtime of the subsequent run. This is reported in Table \ref{tab:basic_intrinsics_subsequent}, where \emph{total} is the total runtime, \emph{xfer} is the data transfer component and \emph{comp} is the compute time component. It can be seen that the runtime of subsequent runs on the NPU is significantly smaller than the initial run, is competitive with, and sometimes outperforms, the CPU. It can also be seen that int32 and float32 are slower on the NPU than their int16 and bfloat16 counterparts, because the former are not supported directly in hardware by the AIE-ML.

\begin{table*}[htb]
    \begin{center}   
    \begin{tabular}{|c|ccc|ccc|ccc|ccc|}
    \hline     
      \textbf{Intrinsic} & \multicolumn{3}{c|}{\makecell{\textbf{int16} \\ \textbf{\textit{runtime (us)}}}} & \multicolumn{3}{c|}{\makecell{\textbf{int32} \\ \textbf{\textit{runtime (us)}}}} & \multicolumn{3}{c|}{\makecell{\textbf{bfloat16} \\ \textbf{\textit{runtime (us)}}}} & \multicolumn{3}{c|}{\makecell{\textbf{float32} \\ \textbf{\textit{runtime (us)}}}}\\     
     & \textbf{total} & \textbf{xfer} & \textbf{comp} & \textbf{total} & \textbf{xfer} & \textbf{comp} & \textbf{total} & \textbf{xfer} & \textbf{comp} & \textbf{total} & \textbf{xfer} & \textbf{comp}\\
      \hline
    sum & 221 & 26 & 195 & 373 & 77 & 296 & 312 & 28 & 284 & 395 & 81 & 314\\
    product & 232 & 28 & 204 & 384 & 74 & 310 & 324 & 27 & 297 & 397 & 79 & 318\\
    maxval & 314 & 28 & 286 & 333 & 79 & 254 & 328 & 28 & 300 & 306 & 72 & 234\\
    minval & 312 & 28 & 284 & 327 & 78 & 249 & 335 & 28 & 307 & 316 & 74 & 242\\
    \hline
    \end{tabular}
     \caption{Runtime performance (in microseconds) of reduction based Fortran intrinsics for subsequent runs on the NPU's AIE array, operating on a one dimensional array of size 262144 elements. \emph{xfer} is data transfer time, and \emph{comp} compute time.}
    \label{tab:basic_intrinsics_subsequent}
    \end{center}
    \vspace*{-\baselineskip}
\end{table*}

We then explored offloading the \emph{transpose} Fortran intrinsic onto the AIE array by leveraging AMD's \emph{transpose-dma} example \cite{mlir-aie}. This example requires no involvement from the compute itself, but instead leverages the compute tile's data mover to undertake the transposition. We modified the code to instead use a memory tile, providing 512KB instead of 64KB in the compute tile. Given the memory tile's fast memory and dedicated data movers, resulting in 30 GB/s bandwidth, it was our hypothesis that this could be beneficial compared to the CPU. Table \ref{tab:transpose} reports a performance comparison of undertaking transposition on the NPU's AIE array against the CPU for the \emph{int32} data type. We report both the first execution runtime on the NPU and subsequent runtimes. Performance on the NPU is fairly flat regardless of the array size, and in comparison performance grows in line with the data size on the CPU. For the largest array that can fit within the memory tile, the NPU outperforms the CPU, however the memory tile's 512KB memory tile is a significant limitation to the size of array that can be handled. Regardless, this again demonstrates the seamless use of the NPU without any modifications required to the Fortran code.

\begin{table}[htb]
    \begin{center}  
    \begin{tabular}{|c|c|c|c|}
    \hline     
      \makecell{\textbf{Array} \\ \textbf{size}} & \makecell{\textbf{CPU} \\ \textbf{runtime (us)}} & \makecell{\textbf{NPU first} \\ \textbf{runtime (us)}} & \makecell{\textbf{NPU subsequent} \\ \textbf{runtime (us)}}\\
      \hline
    64x64 & 11 & 358 & 194 \\
    128x128 & 151 & 446 & 235\\
    256x256 & 203 & 418 & 240 \\
    512x256 & 576 & 440 & 230 \\
    \hline
    \end{tabular}
    \caption{\emph{transpose} Fortran intrinsic runtime with int32}
    \label{tab:transpose}
    \end{center}
    \vspace*{-\baselineskip}
\end{table}

The intrinsics considered so far in this section, although commonplace in Fortran codes, have been rather simple computationally. The Fortran standard defines a matrix multiplication, \emph{matmul}, intrinsic which is much more computationally intensive and has also been optimised on the AIEs by AMD. We leveraged AMD's matrix multiplication example \cite{mlir-aie} that runs across the sixteen AIEs of the NPU, and integrated this with our compilation flow. Consequently, when the programmer calls the Fortran \emph{matmul} intrinsic from their code this is then launched on the NPU.

\begin{table}[htb]
    \begin{center}    
    \begin{tabular}{|c|c|c|c|}
    \hline     
      \makecell{\textbf{Data} \\ \textbf{type}} & \makecell{\textbf{CPU} \\ \textbf{runtime (us)}} & \makecell{\textbf{NPU first} \\ \textbf{runtime (us)}} & \makecell{\textbf{NPU subsequent} \\ \textbf{runtime (us)}}\\
      \hline
    int16 & 5473 & 2572 & 1353 \\
    int32 & 14032 & 2635* & 1503*\\
    bfloat16 & 815194 & 2626 & 1357 \\
    float32 & 17566 & 3901* & 1471* \\
    \hline
    \end{tabular}
    \caption{Runtime (in microseconds) of \emph{matmul} Fortran intrinsic with a problem size of 256x256x512 elements}
    \label{tab:matrix_mul}
    \end{center}
    \vspace*{-\baselineskip}
\end{table}

Table \ref{tab:matrix_mul} reports the performance of the matrix multiplication Fortran intrinsic for input input array sizes of 256x256 and 256x512, calculating a result array size of 256x512, across the CPU and NPU for different data types. It can be seen that the NPU outperforms the CPU for all configurations. The int32 and float32 calculations on the NPU are marked with an asterisk because this data type is used as the output, with the algorithm using the reduced precision counterpart (int16 and bfloat16 respectively) for inputs. The matrix multiplication kernel running on the NPU has been heavily optimised by AMD, taking advantage of vectorised multiply accumulate operations, and this suits the AIE architecture. This demonstrate the benefits of running suitable operations on the NPU, and although the AIE code itself is complicated and highly specialised this is all hidden from the Fortran programmer.

\section{Conclusions and further work}
\label{sec:conclusions}

Whilst the specialised computation provided by AMD's AI engines has the potential to deliver improved performance for appropriate workloads, especially as now these are integrated with AMD CPUs, it is not realistic to expect scientific programmers to have the required architecture specific knowledge and expertise. Consequently, in this paper we have described an approach which enables seamless offloading of Fortran intrinsic calls onto AMD's Ryzen AI AIE array. Building upon AMD's MLIR AIE support, we developed an \emph{xrt\_wrapper} MLIR dialect to drive the NPU from the CPU, and generate IR for the AIEs based upon templates that stored in a library that are specialised by our approach for each instantiation.

We demonstrated that, whilst there is some initial setup overhead associated with running on the NPU, if intrinsics are repeatedly called from Fortran code, which is common in scientific computing, then this overhead can be ameliorated. This is especially important for fairly simple reduction based intrinsics, whereas the specialised nature of the NPU provides more clear benefits for array transposition at larger data sizes and especially for matrix multiplication. 

In this paper we have focused on Fortran due to its popularity in scientific computing, but the linear algebra dialect is also leveraged by the ONNX dialect which is used extensively by ML frameworks, and future work will be to explore coupling and optimising our approach with those frameworks. Whilst we have focused on intrinsics in this paper, it would also be interesting to extend this work to a wider range of algorithmic patterns such as stencils. Work has already been undertaken mapping an MLIR stencil dialect to FPGAs \cite{bisbas2024shared}, and we plan on extending this to also target AIEs.

We conclude that Fortran intrinsics can be offloaded to the NPU in the AMD Ryzen AI CPU by the compiler, enabling programmers to trivially leverage the AIE's specialised compute without expret knowledge or effort required on their behalf.


\bibliographystyle{ACM-Reference-Format}
\bibliography{sample-base}

\end{document}